%% file: ProtonsPbPaper.tex
\documentclass[ALICE,manyauthors]{cernphprep}
\usepackage[comma,square,numbers,sort&compress]{natbib}

\usepackage{lineno}
\usepackage{xspace}
\usepackage{hyperref}
\usepackage{multirow}
\usepackage[dvipsnames]{xcolor}
\usepackage[T1]{fontenc}
\usepackage{orcidlink}

\begin{document}
\input{commands.tex}

\begin{titlepage}
\PHyear{2024}       
\PHnumber{288}      
\PHdate{06 November}  

\title{Proton emission in ultraperipheral Pb--Pb~collisions at $\mathbf{\sqrt{\textit{s}_{\mathrm{\textbf{NN}}}}=5.02}$~TeV }
\ShortTitle{Proton emission in UPC Pb–Pb at $\sqrt{s_{\mathrm{NN}}}=5.02$~TeV}   

\Collaboration{ALICE Collaboration\thanks{See Appendix~\ref{app:collab} for the list of collaboration members}}
\ShortAuthor{ALICE Collaboration} 

\begin{abstract}

The first measurements of proton emission accompanied by neutron emission in the electromagnetic dissociation (EMD) of  $^{208}$Pb nuclei in the ALICE experiment at the LHC are presented. The EMD protons and neutrons emitted at very forward rapidities are detected by the proton and neutron Zero Degree Calorimeters of the ALICE experiment. The emission cross sections of zero, one, two, and three protons accompanied by at least one neutron were measured  in ultraperipheral $^{208}$Pb--$^{208}$Pb collisions at a center-of-mass energy per nucleon pair $\sqrt{s_{\rm NN}}=5.02$~TeV. The 0p and 3p cross sections are described by the RELDIS model within their measurement uncertainties, while the 1p and 2p cross sections are underestimated by the model by 17–25\%. According to this model, these 0p, 1p, 2p, and 3p cross sections are associated, respectively, with the production of various isotopes of Pb, Tl, Hg, and Au in the EMD of $^{208}$Pb. The cross sections of the emission of a single proton accompanied by the emission of one, two, or three neutrons in EMD were also measured. The data are significantly overestimated by the RELDIS model, which predicts that the (1p,1n), (1p,2n), and (1p,3n) cross sections are very similar to the cross sections for the production of the thallium isotopes $^{206,205,204}$Tl.

\end{abstract}
\end{titlepage}

\setcounter{page}{2} 


\section{Introduction}

Ultraperipheral collisions (UPCs) of relativistic (heavy) ions are characterized by no geometric overlap of nuclear volumes. In UPCs, there are no short-range hadronic interactions between colliding nuclei, but either nucleus can still be excited by the long-range Lorentz-contracted electromagnetic field of its collision partner. In such collisions electromagnetic forces can cause the electromagnetic dissociation (EMD) of nuclei~\cite{Pshenichnov2011}. The dominant process of the emission of one and two neutrons from heavy nuclei in EMD is induced mostly by the absorption of low-energy Weizs\"{a}cker--Williams photons~\cite{Bertulani1987} which dominate in UPCs.  This leads to the excitation and decay of Giant Dipole Resonances (GDR) in the colliding nuclei below the proton emission threshold.  When photons of higher energy are absorbed by heavy nuclei in UPCs, several neutrons are expected to be emitted, possibly accompanied by protons.

The cross sections of the emission of 1, 2, 3, 4, and 5 forward neutrons in EMD with and without forward protons were recently measured by ALICE at a center-of-mass energy per nucleon pair $\sqrt{s_{\rm NN}} = $~5.02~TeV in UPCs of $^{208}$Pb~\cite{ALICE2023n}, complementing and  extending previous ALICE measurements at $\sqrt{s_{\rm NN}} = $~2.76~TeV~\cite{Abelev2012n}, and compared with the RELDIS~\cite{Pshenichnov2011} and $\mathrm{n^O_On}$~\cite{Broz2020} model predictions. Data and models were found to be in good agreement for the 1n and 2n cross sections, but differed for higher neutron multiplicities. The measurements~\cite{ALICE2023n} have provided valuable input for the validation of the RELDIS~\cite{Pshenichnov2011}, $\mathrm{n^O_On}$~\cite{Broz2020}, and other  models~\cite{Braun2014,Klusek-Gawenda2014} aimed at describing EMD. 

The sums of the measured cross sections for the emission from 1 to 5 neutrons (1n--5n) in ultraperipheral  $^{208}$Pb-- $^{208}$Pb collisions at $\sqrt{s_{\rm NN}} = $~5.02~TeV with an arbitrary number of protons (including zero protons) and without protons (0p) amount, respectively, to $151.5\pm 4.7$~b and $126.0\pm 4.1$~b~\cite{ALICE2023n}. Both values represent the dominant contributions to the total EMD cross section of $206.5 \pm 5.0$~b, estimated as $(\sigma_{\mathrm{ZED}}-\sigma_{\mathrm{had}})/2$. Here $\sigma_{\mathrm{ZED}}$ is defined as the cross section measured with the ZED trigger requiring at least one neutron to be detected on either side of the interaction point (IP), and $\sigma_{\mathrm{had}}$ is the total hadronic cross sections measured in Ref.~\cite{Castellanos2021}. In this estimation  a 100\% efficiency of the ZED trigger is assumed. From the comparison of the two aforementioned sums it can be concluded that about 17\% of the 1n--5n EMD events include proton emission. However, less is known about the EMD of $^{208}$Pb at the LHC induced by more energetic photons and represented by the emission of protons along with higher numbers of neutrons. According to theoretical estimates with the RELDIS~\cite{ALICE2023n, Dmitrieva2023} and FLUKA~\cite{Braun2014} models, this leads to the production of thallium (Tl), mercury (Hg), gold (Au), and other elements as secondary nuclei and contributes significantly to the decay of the beam intensity~\cite{Baltz1996}. These secondary nuclei cannot be identified and directly measured at the LHC. In contrast, at the CERN SPS the production of secondary nuclei in the fragmentation of $^{208}$Pb in collisions with Au and Pb targets were directly measured by two experiments~\cite{Scheidenberger2004,Cecchini2002} at  $\sqrt{s_{\mathrm{NN}}}=17.21$~GeV. However, neither of them  were geared to detect forward neutrons or protons. 

Furthermore, studies of the proton emission at the LHC can also provide insights into photon-induced reactions for the modeling and design of the upcoming Electron--Ion Collider (EIC)~\cite{EIC_CDR}, where precision measurements of nucleon structure and parton dynamics in high-energy electron--ion collisions are a central focus. For instance, these measurements can help to refine theoretical models and experimental techniques for the study of $J/\psi$ photoproduction, where it is important to know the proton and neutron emission yields in order to reject incoherent background~\cite{Chang2021}. 

RELDIS predicts~\cite{Dmitrieva2023} that the cross sections of the production of secondary nuclei, in particular, thallium, mercury, and gold,  can be well approximated by the cross sections of the emission of the corresponding numbers of protons and neutrons. The ALICE experiment is equipped with both neutron and  proton  Zero  Degree Calorimeters (ZDCs): the ZN for the detection of neutrons and the ZP for the detection of protons. This combination of detectors  enables the  study of EMD leading to the emission of forward neutrons accompanied by forward protons and vice versa.

This work reports the measured cross section of proton emission in EMD in ultraperipheral  $^{208}$Pb-- $^{208}$Pb collisions at $\sqrt{s_{\rm NN}} = $~5.02~TeV, namely, of one, two, and three protons (1p, 2p, and 3p) accompanied by at least one neutron. The cross sections of the emission of a single proton accompanied by one, two, and three neutrons, (1p,1n), (1p,2n), and (1p,3n), are also presented. These cross sections are associated, respectively, with the production of Tl, Hg, and Au nuclei, and with the production of the isotopes: $^{206,205,204}$Tl. The 0p cross section, associated with the production of various Pb isotopes as secondary nuclei, was also measured. The measurements are compared with RELDIS results and previous measurements at lower energies~\cite{Scheidenberger2004,Cecchini2002}. 

The paper is organized as follows. In Sec.~\ref{Sec:Setup} the ALICE ZDCs are briefly described. In Sec.~\ref{Sec:ZED} the data sample and the adopted trigger configuration are presented. In Sec.~\ref{Sec:Analysis} the methods of fitting the ZDC energy spectra are described. The corrections for detection efficiency and acceptance used to obtain the measured cross sections are presented in Sec.~\ref{Sec:Results}, where also the difference between measured and calculated cross sections is discussed.  Finally, in Sec.~\ref{Sec:Conclusions} conclusions and an outlook are given.

\section{Experimental setup}\label{Sec:Setup}

A detailed description of the ALICE experiment and its performance can be found in Refs.~\cite{ALICE2008JINST,ALICE2014performance}. The ALICE setup is equipped with two identical systems of forward hadronic calorimeters (ZDCs) placed on both sides (C and A) of the IP, 112.5~m away from the nominal IP position along the beam direction. The ZDCs are employed to detect the energy carried away in the forward direction by spectator nucleons, thereby enabling the measurement of the number of spectator nucleons in the collision. The neutron (ZNC and ZNA) and proton (ZPC and ZPA) calorimeters, see Refs.~\cite{Puddu2007,Gemme2009,Oppedisano2009} for details, are denoted as ZN and ZP, respectively. ZNC and ZNA are placed at zero degrees with respect to the LHC beam axis to detect neutral forward particles at pseudorapidities $|\eta| > 8.8$. ZPC and ZPA  are positioned externally to the outgoing beam vacuum tube and serve to detect forward protons with an energy close to that of the beam directed into these calorimeters by the LHC magnet system. The ZDCs are supplemented by two small electromagnetic calorimeters (ZEM1 and ZEM2) placed only on the side A, at 7.5 m from the IP, covering the pseudorapidity range $4.8 < \eta < 5.7$ and about 18\% of the total $2\pi$  azimuthal angle in the two intervals  $-16^\circ<\phi<16^\circ$ and $164^\circ<\phi <196^\circ$~\cite{Oppedisano2009}. Signals are expected in ZEM1 or ZEM2 in hadronic $^{208}$Pb-- $^{208}$Pb events with high particle multiplicity in a wide pseudorapidity range, but not in EMD events represented only by low multiplicity of very forward particles. Therefore, the ZEM veto is very effective in selecting EMD events, as shown in Ref.~\cite{ALICE2023n}.  The detailed technical characteristics of the ZN, ZP, and ZEM calorimeters used also in the present measurements are given in Ref.~\cite{ALICE2023n}.

\section{Data sample}\label{Sec:ZED}

The present analysis is based on a data set of ultraperipheral $^{208}$Pb--$^{208}$Pb collisions collected by ALICE in 2018, and which was recently used to measure the cross sections of 1n--5n emission in EMD of $^{208}$Pb at $\sqrt{s_{\rm NN}} = $~5.02~TeV~\cite{ALICE2023n}. In special runs for the EMD cross section measurements, the average number of hadronic inelastic interactions per bunch crossing $\mu_{\mathrm{inel}} \sim 1.3\times 10^{-4}$ was about 10 times lower than during the ALICE standard physics runs due to reduced instantaneous luminosity. Thus, the event pileup was also reduced. The events were triggered when a signal exceeded a threshold of about 500 GeV in ZNC or ZNA. This defines the ZED trigger condition. This trigger is sensitive to the single EMD events with neutrons either on the C or A side, as well as to mutual EMD and hadronic processes with neutrons detected on both sides. Using the ZDC timing information, only events corresponding to interactions from the nominal bunch crossing were selected, thus rejecting beam--gas interactions and collisions between nuclei outside the nominal bunch crossings. The total number of selected ZED events is ${N_{\mathrm{tot}}} = 2.050 \times 10^{6}$. Luminosity normalization is performed via the ZED trigger cross section $\sigma_{\mathrm{ZED}}= 420.6 \pm 10.1$~b~\cite{Castellanos2021} measured in a van der Meer (vdM) scan~\cite{vanderMeer1968}.

The condition of absence of signals in both ZEM1 and ZEM2 allows electromagnetic events to be distinguished from hadronic ones~\cite{Cortese2019}. The very limited azimuthal angle and pseudorapidity coverage of the ZEMs make the rejection of low-multiplicity EMD events by  the ZEM veto very unlikely. According to RELDIS~\cite{Pshenichnov2011}, the average total  number of charged particles in an EMD event is equal to 1.55 and only $\sim 0.03$ of them are emitted within the ZEM acceptance in $^{208}$Pb--$^{208}$Pb collisions at $\sqrt{s_{\rm NN}} = $~5.02~TeV. As reported in Ref.~\cite{ALICE2023n}, the efficiency of selecting EMD events by imposing the ZEM veto remains extremely high ($>99.8$\%) for the emission of 1–5 neutrons. 

As shown by previous ALICE measurements~\cite{ALICE2023n}, no protons (0p) are emitted in a dominant fraction ($\sim 83$\%) of 1n--5n EMD events. Due to the overlap of 1n--5n EMD events with 0p events, the reported high efficiency of the ZEM veto can be safely extrapolated to 0p emission measured in the present work. The high efficiency of the ZEM veto in discriminating 1p, 2p, 3p and (1p,1n), (1p,2n), (1p,3n) EMD events from hadronic ones was additionally confirmed by calculations with the RELDIS and AAMCC-MST~\cite{Nepeivoda2022,Kozyrev2022} models.

It was estimated with RELDIS that only 1.3\%, 1.0\%, 0.8\% of 1p, 2p, 3p and 1.7\%, 1.5\%, 1.2\% of (1p,1n), (1p,2n), (1p,3n) events, respectively, are rejected by the ZEM veto. However, peripheral hadronic $^{208}$Pb--$^{208}$Pb collisions with the same number of forward nucleons, emitted as spectators, can potentially mimic EMD events if they are not rejected by the ZEM veto. In peripheral hadronic events of the  80\% to 90\% centrality class, $l = 15.4\pm 2.1$ charged particles are emitted on average in the pseudorapidity interval $4<\eta<5$~\cite{ALICE2023s} with a very flat  $\eta$ distribution. The values of ${\rm d}N_{\rm ch}/{\rm d}\eta$ measured for the 80-90\% centrality class can be extrapolated to the last centrality class of 90–100\% based on the Monte Carlo Glauber model~\cite{Loizides2017}. According to this model, the average number of inelastic NN collisions is about 2.5 times lower when moving to the next centrality class in the row: 70--80\%, 80–90\%, 90–100\%. This gives  a number of charged particles $l\approx 6.16$ in  $4<\eta<5$ for the 90–100\% centrality class, and $l\approx 10.8$ on average in the 80-100\% centrality class. Given the flatness of ${\rm d}N_{\rm ch}/{\rm d}\eta$, it can be expected that about $l=10$ particles are emitted on average in the interval $4.8<\eta<5.7$ covered by ZEMs. 
However, due to the very limited angular acceptance of ZEMs most of the particles do not produce ZEM signals. A simple estimate assuming a random emission of each of these particles within $2\pi$ of the azimuthal angle gives a probability $p=0.18$ of hitting the ZEM. Therefore, the probability that all $l$ particles in a peripheral hadronic event will miss the ZEM is $(1-p)^l=0.137$. One can consider this value as a conservative estimate because a lower probability of 0.076 for hadronic events to survive the ZEM veto has been calculated previously by Monte Carlo simulations with the HIJING model for lower collision energy~\cite{Abelev2012n}. The corresponding cross section for peripheral (80--100\%) hadronic events potentially contaminating EMD events is rather small: 1.54~b$\times 0.137 = 0.21$~b and it is shared between several partial EMD cross sections making it negligible in each particular EMD channel.

In the present study, the AAMCC-MST~\cite{Nepeivoda2022,Kozyrev2022} model was used to calculate the cross sections of the emission of a given number of spectator nucleons in  hadronic $^{208}$Pb--$^{208}$Pb collisions. Finally, taking into account the ZEM veto factor of 0.137, the hadronic contamination of EMD events was estimated to be 0.3\%, 0.9\%, 1.3\% for 1p, 2p, 3p and 0.03\%, 0.09\%, 0.16\% for (1p,1n), (1p,2n), (1p,3n), respectively, which can be neglected given other uncertainties of the EMD cross section measurements listed in Table~\ref{tab:syst_uncert}.
Following this consideration, events with a signal in either ZEM were considered as hadronic and rejected, yielding a sample of ${N^{\mathrm{EMD}}}_{\mathrm{tot}} = 2.009 \times 10^{6}$ single or mutual EMD event candidates.

\section{Analysis}\label{Sec:Analysis}

The methods used to collect, calibrate, and fit the ZDC energy spectra follow basically the same steps as in Ref.~\cite{ALICE2023n}. For this reason, only analysis features specifically related to the emission of protons in EMD are described in the following.

In Ref.~\cite{Dmitrieva2023} the distributions of the total energy of neutrons and protons from EMD of $^{208}$Pb at $\sqrt{s_{\rm NN}} = $ 5.02~TeV were calculated without accounting for the energy resolution and efficiency of the ZDCs. It was found that such pristine distributions of $E_{\mathrm{ZN}}$ and $E_{\mathrm{ZP}}$ are very different because of the difference in the energies of emitted neutrons and protons in the rest frame of $^{208}$Pb. Indeed, the peaks corresponding to 1n, 2n,~...~7n emission are narrow since the neutrons are mainly evaporated subsequently from an excited residual nucleus, and the energy of each neutron is typically below 1--2 MeV in the nucleus rest frame. In contrast, the proton peaks are much wider, since these protons are mainly produced by photons with $E_\gamma > 140$~MeV and thus, share a noticeable part of the photon energy along with photoproduced pions~\cite{Dmitrieva2023}. As explained in Ref.~\cite{Pshenichnov1999}, since a larger phase space is available for multiply produced secondary particles, some protons and pions are emitted backward or forward with respect to the initial photon momentum. When the proton momenta are Lorentz-boosted to the laboratory system, the deviation of proton energies from the beam energy of $E_{0}=2.51$~TeV can  be  significant,  depending also on the direction of their emission. This explains the much wider proton peaks compared to the neutron peaks and also indicates the difficulty in measuring  proton EMD cross sections.

\subsection{Fit of the ZP energy spectra}\label{Sec:ZPSpectra}

The number of events, $n_k$ and $n_i$, detected for each proton $k$p and neutron $i$n multiplicity class,  were extracted by fitting the calibrated distributions of the energy $E$ deposited in the ZP or ZN, respectively. The $\chi^2$ fit method was used. The procedures to fit the measured energy distributions in the ZN with the sum of Gaussians representing events of each neutron  multiplicity were validated previously with Run~1~\cite{Abelev2012n} and with Run~2~\cite{ALICE2023n} data. 

\begin{figure}[t]
\begin{centering}
\includegraphics[width=1.0\columnwidth]{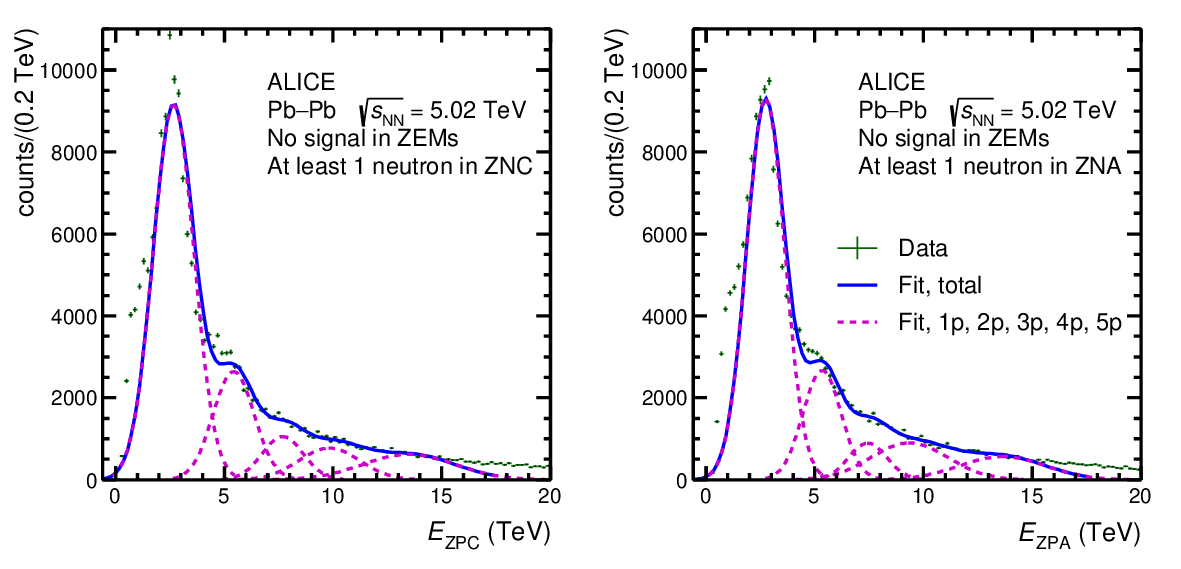}
\caption{Distributions of the energy  measured in ZPC (left) and ZPA (right) from the EMD events  with at least one neutron at the same side (points) and resulting fit functions (solid curves) representing the sum of the individual Gaussians. The Gaussians representing the contributions of the individual 1p, 2p, 3p, 4p, and 5p peaks are shown by the dashed curves.
}
\label{fig:Spectr_EM_p}
\end{centering}
\end{figure}

The distributions of the energy deposited by EMD protons in ZPC and ZPA are shown in Fig.~\ref{fig:Spectr_EM_p}. These distributions represent events which have at least one neutron detected in the neutron ZDC on the same side where the protons are registered.  This means that the EMD events with protons, but without neutron emission, are not included in these measurements. According to the RELDIS model~\cite{Pshenichnov2011}, such neutronless EMD events account for about 3\% of the total EMD events. 

As can be seen in Fig.~\ref{fig:Spectr_EM_p}, the low heights and large widths of the peaks corresponding to $k>2$ protons in the ZP energy spectra make it difficult to determine the number of events with more than two protons. Therefore, the fitting procedure previously used to fit the energy distributions in ZN~\cite{ALICE2023n,Abelev2012n} with the narrower 3n, 4n, and 5n peaks had to be adapted to deal with the present ZP spectra. A modified fitting function $\widetilde{F}(E)$ representing the sum of $j$ unnormalized Gaussians was used in the present analysis to reduce the correlation between the parameters describing the height and width of each peak:
\begin{equation}
\widetilde{F}(E)=\sum_{k=1}^{j} \widetilde{G_k}(E) =\sum_{k=1}^{j} \widetilde{n_k} e^{-\frac{(E-\mu_k)^2}{2\sigma_k^2}} \ .
\label{Eq:FitFunc_u}
\end{equation}

The Gaussian $\widetilde{G_k}(E)$ represents the peak corresponding to the emission of $k$ protons, and is characterized by the mean value $\mu_k$, the dispersion $\sigma_k$, and the normalization constant $\widetilde{n_k}$, which is related to the number of events $n_k$ in each peak by the equation $n_k = \sqrt{2\pi} \sigma_k \widetilde{n_k}$. The adopted modified fitting procedure consisted of two steps.  At the first step, the 1p peak was fitted individually with $\mu_1$, $\sigma_1$, and $\widetilde{n_1}$ considered as free parameters of the fit with a single Gaussian applied over the energy range from 0 to $2\times E_{0}$, with $E_{0}=2.51$~TeV defined as the $^{208}$Pb beam energy per nucleon. Despite the expected exact correspondence of $\mu_1$ to $E_0$, $\mu_1$ was allowed to vary within the range of $E_0 \pm 20$\%. In addition, the value of $\sigma_1$ was constrained to be between $0.1\times E_0$ and $0.5\times E_0$. These conditions account for a possible deviation of the peak position from $E_0$. At the second step, the ZP spectra were fitted with the sum of the five Gaussians given by Eq.~(\ref{Eq:FitFunc_u}) defined in the energy range from  $-E_{0}/3$ to $(j+2) E_{0}$, with $j=5$, with the fixed values of  $\mu_1$ and $\sigma_1$ taken from the previous step. The ranges of the values of the fit parameters of the Gaussians describing the 2p, 3p, 4p, and 5p peaks were also restricted: $\mu_k$ was allowed to vary within $\pm 20$\% around $k \mu_1$, while $\sigma_k$ was constrained between $\sigma_1$ and $\sqrt{k} \sigma_1$.  The expected values of $\mu_k = k \mu_1$ and $\sigma_k = \sqrt{k} \sigma_1$ were taken to start the fitting procedure. The values of  $\widetilde{n_k}$ in each peak were considered as free parameters of the fit. The resulting  normalized Gaussians representing contributions of the individual 1p, 2p, 3p, 4p, and 5p peaks are shown in Fig.~\ref{fig:Spectr_EM_p}, along with the fit representing their sum. It should be noted, however, that the final fit does not aim to describe the ZP spectra in the region of the 5p peak and above, where the contribution of events with more than five protons is dominant. As described below in Sec.~\ref{Sec:CrossCalc}, after taking into account the acceptance$\times$efficiency of the ZDCs, the true numbers of events $N_k$ were finally obtained from ${n_k}$.

The individual 1p and 2p peaks in the distributions of energy in the ZP shown in Fig.~\ref{fig:Spectr_EM_p} are much wider compared to the 1n and 2n peaks in the distributions of energy in ZN reported in Ref.~\cite{ALICE2023n}. For comparison, the width of the 1n peak is about 0.5 TeV, while the width of the 1p peak approaches a value of 0.9 TeV, which is nearly a factor of two larger than that observed for 1n. The energy resolution of the ZN is only slightly better than the resolution of the ZP according to the extrapolation from test beam data~\cite{Gemme2006}. As discussed above, see also Ref.~\cite{Dmitrieva2023}, the main difference in the widths of the neutron and proton peaks is due to the larger phase space available for protons emitted in EMD compared to neutrons. As a result, the uncertainties of the fit parameters obtained for wide overlapping 
1p, 2p and 3p peaks in the ZP energy spectra are much larger compared to the uncertainties of the parameters of the 1n, 2n, 3n peaks~\cite{ALICE2023n}.       

The shoulders adjacent to 1p peaks visible in Fig.~\ref{fig:Spectr_EM_p} at both the A and C sides can be attributed to those protons which hit the calorimeters near their edges and thus do not deposit their energy entirely inside the ZPs. In addition, protons which lose a part of their energy before reaching the ZP, e.g., in the interactions with the beam vacuum tube, also contribute to this shoulder. This was confirmed by the simulated distributions of $E_{\mathrm{ZP}}$, obtained in a Monte Carlo simulation with protons and neutrons generated with RELDIS~\cite{Pshenichnov2011}, where some shoulders adjacent to pristine 1p peaks were also observed, due to the protons which do not release their entire energy inside the calorimeter. This effect is taken into account in the calculation of the efficiency correction factors.

The final fit is below the data points near the maximum of the 1p peak, while it is above the points to the right of the maximum. As a result, the integrals of the fit function and the data over these two regions differ by less than the systematic uncertainty of the fit.

\subsection{Fit of the ZN energy spectra}\label{Sec:ZNSpectra}

A two-step procedure was implemented to obtain the distributions of energy deposited in neutron ZDCs for events in which a single proton was also detected on the same side. First, two-dimensional correlations between the energies deposited in proton and neutron ZDCs were considered to define an interval of the ZP energy attributed to the emission of a single proton. Second, the events within the selected interval of the ZP energy were projected onto the ZN energy axis to obtain the energy distributions in the ZN with the condition of detecting a single proton in the ZP. 

\begin{figure}[t]
\begin{centering}
\includegraphics[width=1.0\columnwidth]{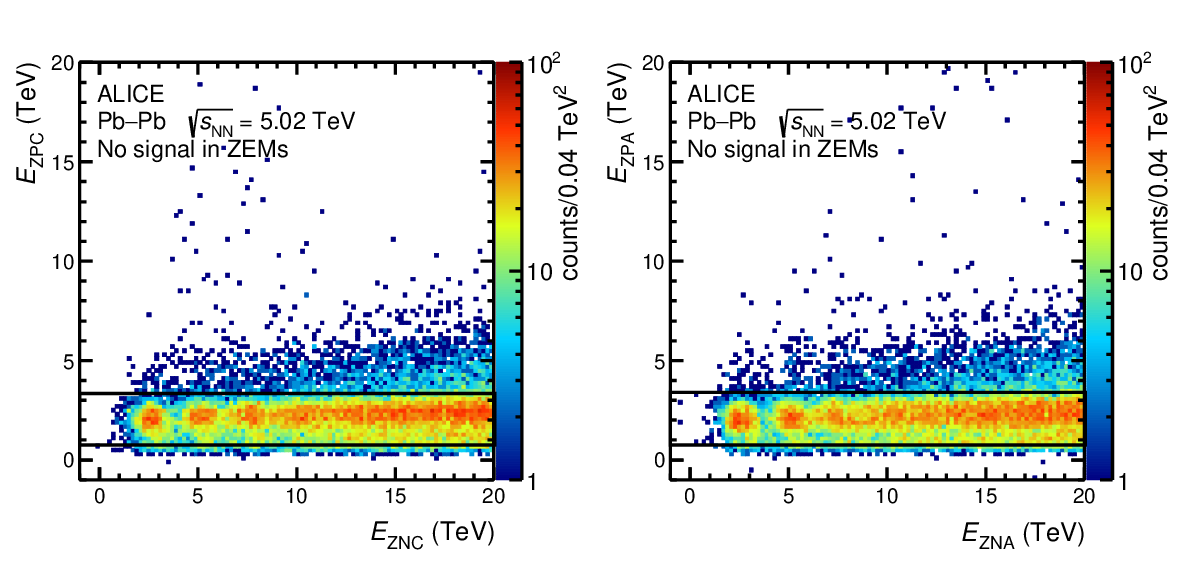}
\caption{Correlations between energies in the ZP and ZN for the EMD events with the emission of up to 7--8 neutrons: ZPC vs ZNC (left) and ZPA vs ZNA (right). The horizontal black lines mark the range $[\mu_1 - 2\sigma_1, \mu_1 + 2\sigma_1]$  of ZPC and ZPA energies which covers $\sim 95$\% of 1p events.
} 
\label{fig:CorrelationsEM}
\end{centering}
\end{figure}

The correlations between energies deposited  in proton and neutron ZDCs on the side C (ZPC vs ZNC) and on the side A (ZPA vs ZNA) are shown in Fig.~\ref{fig:CorrelationsEM}. These correlations between the energies deposited in the neutron and proton ZDCs placed on the same side reveal three distinct spots corresponding to the 1n, 2n, and 3n emission within a horizontal band corresponding to the 1p emission. In contrast, the spots corresponding to the 4n and 5n emission are smeared out. In order to select specifically the 1n--3n events within the 1p band the energies deposited in the ZN were restricted to $E_{\mathrm {ZNC}}<8.5$~TeV and $E_{\mathrm {ZNA}}<8.5$~TeV. The selected data are used to determine the position  $\mu_1$ and width $\sigma_1$ of the first proton peak more accurately. The boundaries of the resulting intervals  $[\mu_1 - 2\sigma_1, \mu_1 + 2\sigma_1]$ of the ZP energies associated with the detection of a single proton are indicated as lines in Fig.~\ref{fig:CorrelationsEM}.

\begin{figure}[!htb]
\begin{centering}
\includegraphics[width=1.0\columnwidth]{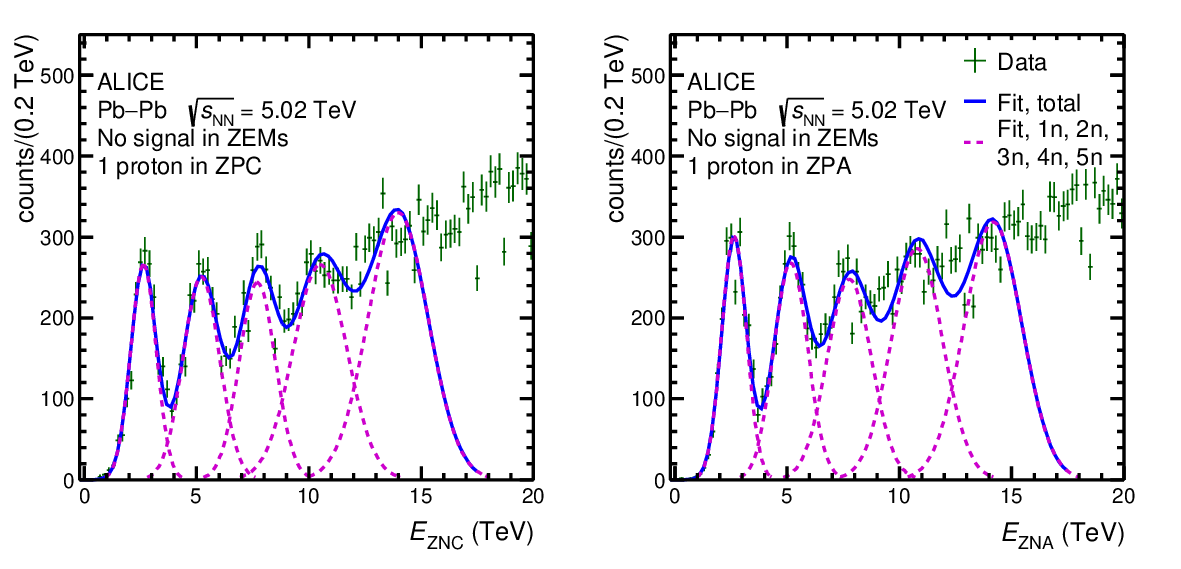}
\caption{Distributions of the energy measured in ZNC (left) and ZNA (right) from the EMD events with a single proton detected, respectively, in ZPC and ZPA (points), and resulting fit functions (solid curves) representing the sum of all the Gaussians. The ZP energies associated with the detection of a single proton are determined as  $[\mu_1 - 2\sigma_1, \mu_1 + 2\sigma_1]$. The Gaussians representing the contributions of individual 1n, 2n, 3n, 4n, and 5n peaks are shown by the dashed curves.
}
\label{fig:Spectr_EM_n_1p}
\end{centering}
\end{figure}

After identifying the domains of 1p events for the C and A sides, the respective energy distributions of the accompanying neutrons in ZNC and ZNA were obtained. These distributions are shown in Fig.~\ref{fig:Spectr_EM_n_1p}, and they demonstrate noticeable statistical fluctuations, in particular, in the regions of the 4n and 5n peaks.   

The procedure used to fit these ZN energy distributions was the same as that used for the ZP energy distributions. However, at the first step $\mu_1$ was constrained to be equal  to $E_0$ within a $\pm 5$\% deviation and the Gaussian was fitted in the energy range from 0 to 1.5$\times E_0$, and at the second step the width $\sigma_1$ of the first neutron peak was allowed to vary within $\pm 20 \%$ instead of keeping it fixed from the previous step. The results of the fit are presented in Fig.~\ref{fig:Spectr_EM_n_1p}.

\subsection{Correction for acceptance and detection efficiency of neutron and proton ZDCs}
\label{Sec:corrections}
Some of the nucleons emitted in the very forward direction are lost along their path to the ZDCs.  In particular, protons scatter on the LHC beam vacuum tubes or lose  energy due to interactions with collider components closer to the ZDC location. In the latter case, the produced secondary particles can reach the calorimeters and contribute to the total energy measured by the ZP. In some events, due to a peripheral impact on the proton calorimeter characterized by additional shower leakage, the energy deposited by the protons is reduced relative to the central incidence. The proton spots on the front surface of the ZP are elongated and extend beyond the lateral edges of the ZP. This is in contrast to the well-centered and focused neutron spots in the neutron ZDCs. However, a very small number of scattered forward neutrons and secondary particles produced by them can also hit the proton ZDC and vice versa. All of these effects alter the distributions of the energy deposited in the ZP and thereby affect the determination of the true proton multiplicity~\cite{Dmitrieva2018}. To obtain the true number of events, the number of events detected for each proton multiplicity have to be multiplied by an appropriate correction factor.

In a realistic Monte Carlo simulation of the ALICE setup, taking into account magnetic fields and the geometry of the beam vacuum tube, after transporting protons and neutrons generated with RELDIS~\cite{Pshenichnov2011}, the number of generated events $N^{\mathrm{MC}}_k$ and the number of events $n^{\mathrm{MC}}_k$ registered in the ZP were calculated for each proton multiplicity class $k$p. The acceptance$\times$efficiency correction factors $f_{k{\mathrm{p}}}=N^{\mathrm{MC}}_k/n^{\mathrm{MC}}_k$ were calculated for ZPC and ZPA, and they are listed in Table~\ref{tab:p_reg_eff}. Due to the fact that protons emitted in EMD can escape detection, events of a given true proton multiplicity are registered as events of lower detected proton multiplicity. This migration towards lower proton multiplicities leads to a net loss of events of proton multiplicity classes with $k > 0$, which explains why $f_{k{\mathrm{p}}}$ is greater than 1 in these cases (see  Table~\ref{tab:p_reg_eff}).  On the other hand, many of the migrated events end up in the proton multiplicity class 0p, resulting in a net gain in this class and therefore the value of $f_{0{\mathrm{p}}}$ is less than 1.
\begin{table}[b]
\caption{Acceptance$\times$efficiency correction factors for detecting protons in ZPC and ZPA and their estimated uncertainties.}
\label{tab:p_reg_eff}
\begin{center}
\begin{tabular}{|c|c|c|}
\hline
$k$p & \multicolumn{2}{c|} {$f_{k \mathrm{p}}$ } \\
 \cline{2-3}
  & {Side C} & {Side A} \\
\hline
0p  & $0.848 \pm 0.015$ &  $0.852 \pm 0.018$   \\
\hline
1p  & $ 1.700 \pm 0.058 $ & $1.666  \pm 0.078$  \\
\hline
2p  & $ 2.450 \pm 0.173$ & $ 2.405  \pm 0.097$ \\
\hline
3p  & $2.760 \pm 0.653$ & $2.570 \pm 0.752 $\\
\hline 
\end{tabular}
\end{center}
\end{table}

Since the proton emission cross sections were measured under the condition that at least one neutron was detected on the same side, the measurements of ${n_k}$ for protons had to be additionally corrected for the efficiencies to detect an arbitrary activity Xn, with X~$>0$, in the ZNs. The values of the correction factors for Xn are given in Table~\ref{tab:fxn}. 
\begin{table}[t]
\caption{Acceptance$\times$efficiency correction factors and their estimated uncertainties for detecting an arbitrary ZN activity provided that at least one forward neutron is emitted.} 
\label{tab:fxn}
\begin{center}
\begin{tabular}{|l|c|c|}
\hline
& ZNC & ZNA  \\
\hline
${f}_{\mathrm{Xn}}$ & 1.072 $\pm$ 0.019  & 1.074 $\pm$ 0.019  \\
\hline 
\end{tabular}
\end{center}
\end{table}

Similarly, the correction factors $f_{1\mathrm{p},i{\mathrm{n}}}$, $i=1,2,3$ for the 1p emission accompanied by 1--3 neutrons were obtained from the Monte Carlo simulation of the ALICE setup also using the EMD events generated with RELDIS. The resulting efficiency correction factors are listed in Table~\ref{tab:n_reg_eff}. 
\begin{table}[t]
\caption{Acceptance$\times$efficiency correction factors for detecting a single proton on the sides C and A accompanied by the detection of one, two, and three neutrons on the respective side.}
\label{tab:n_reg_eff}
\begin{center}
\begin{tabular}{|c|c|c|c|}
\hline
 $k$p & $i$n   & \multicolumn{2}{c|}{$f_{1\mathrm{p}, i{\mathrm{n}} }$} \\
\cline{3-4}
 &  & {Side C} & {Side A} \\ 
\hline
 1p  & 1n & 2.437 $\pm$ 0.081    & 2.439  $\pm$ 0.082\\
\hline
 1p & 2n & 2.433 $\pm$ 0.206     & 2.481  $\pm$ 0.130	\\
\hline
 1p & 3n & 2.796  $\pm$ 0.124    & 2.523  $\pm$ 0.206 \\
\hline
\end{tabular}
\end{center}
\end{table}

 All  correction factors were obtained with two different methods, and the simple averages were taken as final correction factors. The first method was based on counting nucleons crossing the front surface of each ZDC, while the second method entailed fitting the simulated ZDC energy spectra with a sum of Gaussians. Finally, in both methods the numbers of events of each multiplicity $i$n or $k$p generated by RELDIS were divided by the numbers of detected events to obtain the correction factors. These two methods are considered as random samples from a space of models rather than two extreme cases. Therefore, the difference between the results of the two methods divided by $\sqrt{2}$ was considered as the systematic uncertainty for the final average correction factors. These factors are given in Tables~\ref{tab:p_reg_eff},~\ref{tab:fxn}, and ~\ref{tab:n_reg_eff} together with their combined statistical and systematic uncertainties. 

\subsection{Determination of proton and neutron emission cross sections}
\label{Sec:CrossCalc}

The cross sections for specific EMD channels $\sigma(k{\mathrm{p}},\mathrm{Xn})$ for a given number of protons $k$ and an arbitrary number of neutrons Xn (X$>0$) were measured individually with ZPC and ZPA. After the signal extraction, the corrections for the efficiency of the ZEM veto and for the ZDC acceptance$\times$efficiency were applied by multiplying by the resulting correction factor $F_{k{\mathrm{p},\mathrm{Xn}}}$ 
\begin{equation}
\sigma(k{\mathrm{p}},\mathrm{Xn})= \sigma_{\mathrm{ZED}}\frac{N_k}{N_{\mathrm{tot}}} = \sigma_{\mathrm{ZED}} \frac{n_k}{N_{\mathrm{tot}}} \frac{f_{k{\mathrm{p}}} f_{{\mathrm{Xn}}}}{\varepsilon_\mathrm{Xn}}=\sigma_{\mathrm{ZED}} \frac{n_k}{N_{\mathrm{tot}}}F_{k{\mathrm{p},\mathrm{Xn}}} \ .
\label{Eq:SigChannelp}
\end{equation}

The efficiency of the ZEM veto to select electromagnetic events, $\varepsilon_i$, for each neutron multiplicity $i$n, or for an arbitrary number of neutrons $\mathrm{Xn}$ (excluding zero neutrons), $\varepsilon_\mathrm{Xn}$, were calculated in Ref.~\cite{ALICE2023n}, separately on the sides C and A, and are reported in Table II of Ref.~\cite{ALICE2023n}. The individual efficiencies $\varepsilon_i$, for $i \leq 3$, are found to be between $99.35$\% and $99.90$\%, while  $\varepsilon_\mathrm{Xn}$ is equal to $96.72$\% and $96.12$\% for the C and A sides, respectively.

Similarly, the cross sections for the 1p emission accompanied by 1--3 neutrons were calculated from the detected number of events ${n_i}$ taking into account all efficiency corrections by multiplying by the resulting correction factor $F_{\mathrm{1p}, i{\mathrm{n}}}$

\begin{equation}
\sigma(\mathrm{1p},i{\mathrm{n}})= \sigma_{\mathrm{ZED}}\frac{N_i}{N_{\mathrm{tot}}} = \sigma_{\mathrm{ZED}} \frac{n_i}{N_{\mathrm{tot}}} \frac{f_{1\mathrm{p}, i{\mathrm{n}} }}{\varepsilon_i} = \sigma_{\mathrm{ZED}} \frac{n_i}{N_{\mathrm{tot}}} F_{\mathrm{1p}, i{\mathrm{n}}}\ .
\label{Eq:SigChanneln1p}
\end{equation}

The final cross sections ${\sigma}$ were obtained as the average of the measurements performed on the sides C and A using the event numbers and resulting correction factors $F$ for each side  introduced in Eqs.~(\ref{Eq:SigChannelp}) and~(\ref{Eq:SigChanneln1p})
\begin{equation}
{\sigma}=\frac{\sigma^\mathrm{C}+\sigma^\mathrm{A}}{2}=\sigma_{\mathrm{ZED}}\frac{n^\mathrm{C} F^\mathrm{C} + n^\mathrm{A} F^\mathrm{A}}{2N_{\mathrm{tot}}} \ \ .
\label{Eq:Average}
\end{equation}

The contributions to the relative systematic uncertainties of ${\sigma}$ are summarized in Table~\ref{tab:syst_uncert}. They were propagated from the systematic uncertainties on each side for the cross sections of the 1p, 2p, and 3p emission accompanied by an arbitrary number of neutrons (Xn, excluding zero neutrons) and for the 1p emission accompanied by the 1n, 2n, and 3n emission. The uncertainty of $N_{\mathrm{tot}}$ is negligible due to the large number of ZED trigger events collected. The statistical uncertainties of $n_k$ ($n_i$) result from the uncertainties of the numbers of events found by the fit procedure. Following Ref.~\cite{Castellanos2021}, the relative systematic uncertainty of $\sigma_{\mathrm{ZED}}$ was taken as $2.4$\%, resulting from the vdM scan analysis. The relative difference between the cross sections calculated with two fitting procedures, termed as main and supplementary, divided by $\sqrt{2}$, was taken as the systematic uncertainty of the fit. In the supplementary fitting procedure the condition  $\sigma_{1} \leq \sigma_{k} \leq \sqrt{k} \sigma_{1}$ ($\sigma_{1} \leq \sigma_{i} \leq \sqrt{i} \sigma_{1}$) for $i, k >1$ was released.  The systematic uncertainties considered for $F_{k\mathrm{p},{\mathrm{Xn}}}$ and $F_{i\mathrm{n},1\mathrm{p}}$ are described in Sec.~\ref{Sec:corrections}.

\begin{table}[b]
\caption{Relative systematic uncertainties of the cross sections of the emission of a given number of protons $k=0, 1, 2, 3$  accompanied by an arbitrary number of neutrons (Xn, excluding zero neutrons) and of the cross sections of the 1p emission accompanied by a given number of neutrons $i=1, 2, 3$ in UPCs of $^{208}$Pb nuclei at $\sqrt{s_{\mathrm{NN}}}=5.02$~TeV. Each uncertainty is calculated for the average of the cross sections measured on the sides C and A.}
\label{tab:syst_uncert}
\begin{center}
\begin{tabular}[c]{|c|c|c|c|c|c|c|c|}
\hline
\multirow{3}*{Source}	& \multicolumn{7}{c|}{Relative uncertainty  (\%)} \\
\cline{2-8}
	& {0p} &  \multicolumn{4}{c|}{1p} & {2p} &{3p}  \\
\cline{2-8}
	& Xn & Xn & \rm{1n} & \rm{2n}& \rm{3n} & Xn & Xn \\

\hline
$\sigma_{\mathrm{ZED}}$ determination& \multicolumn{7}{c|}{\multirow{2}*{2.4}} \\ 
 from vdM scan & \multicolumn{7}{c|}{}\\

\hline
Fitting procedure  & - &  0.8 &  2.4   &  10.2  & 31.8  & 21.3   &  25.6 \\
\hline
ZDC+ZEM efficiency    &  1.8 & 3.1   & 2.3   &5.0   & 4.9  & 4.3   & 18.5  \\

\hline
Total                  &  3.0 & 4.0 & 4.1  & 11.6 & 32.3   & 21.9 & 31.7   \\
\hline
\end{tabular}
\end{center}
\end{table}

\section{Results}\label{Sec:Results}

\subsection{Emission of a given number of protons}
\label{Sec:p_mult}
\begin{figure}[!htb]
\begin{centering}
\includegraphics[width=0.9\columnwidth]{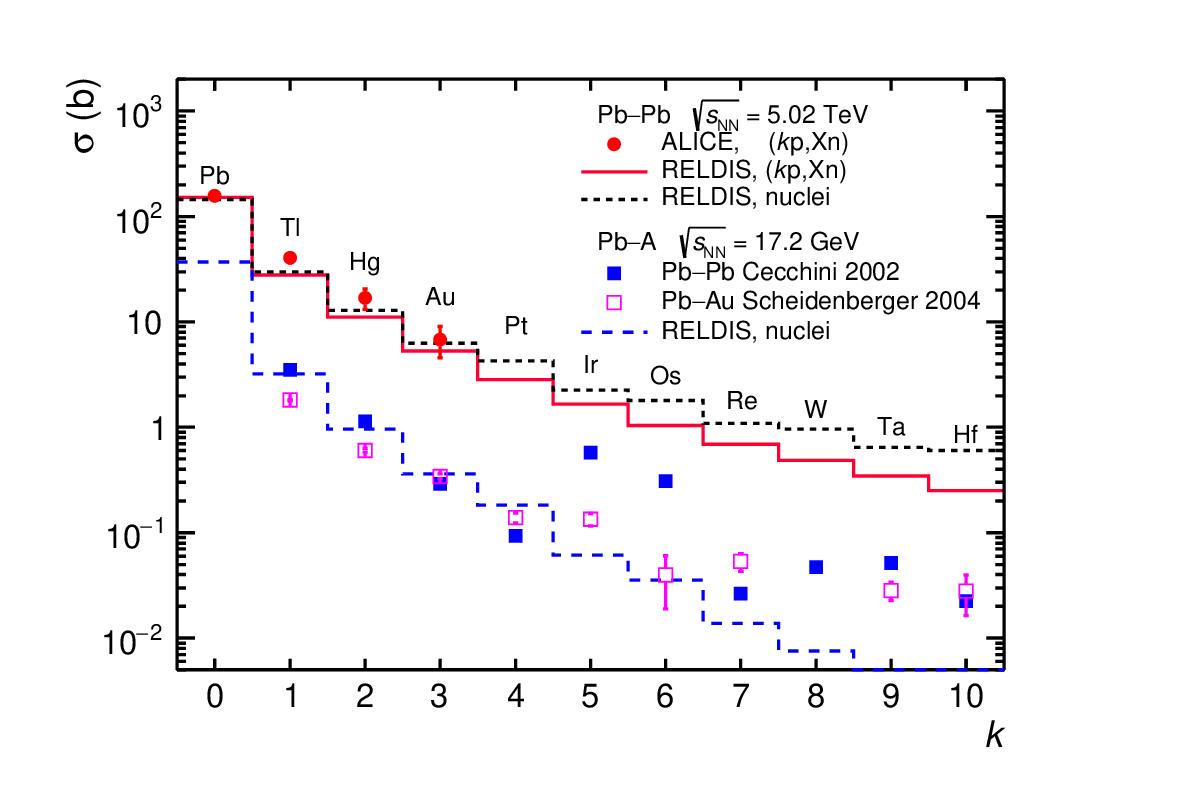}
\caption{Measured (circles) and calculated with RELDIS~\cite{Pshenichnov2011} (red solid-line histogram) cross sections of the emission of a given number of protons $k$ accompanied by at least one neutron in UPCs of $^{208}$Pb nuclei at $\sqrt{s_{\mathrm{NN}}}=5.02$~TeV. Calculated cross sections of the production of specific elements, Pb, Tl, Hg, ..., Hf, at this collision energy are represented by the black dashed-line histograms marked with nuclide symbols.  Charge-changing cross sections of the production of specific elements  measured in collisions of $^{208}$Pb with Au and Pb at $\sqrt{s_{\mathrm{NN}}}=17.21$~GeV  ~\cite{Scheidenberger2004,Cecchini2002} with subtracted hadronic contribution (see text for details) are presented by the solid and open squares and compared with RELDIS results (blue long-dash histogram). The error bars represent combined statistical and systematic uncertainties of the measurements.}  
\label{fig:protons}
\end{centering}
\end{figure}

The measured cross sections of the emission of one, two, and three protons, together with the measured 0p cross section of neutron emission without protons and the sum of 0p--3p cross sections are presented in Table~\ref{tab:p_cs}. The same cross sections calculated with the RELDIS model in the standard mode and without pre-equilibrium emission (see below) are  given for comparison. The measured 0p, 1p, 2p, and 3p cross sections are shown in Fig.~\ref{fig:protons} (red circles), where the theoretical values obtained in the standard RELDIS mode are plotted up to 10p emission (solid red line). As follows from the comparison, the measured and calculated cross sections for the 0p and 3p emission agree with RELDIS within their uncertainties, while the 1p and 2p emissions are underestimated by RELDIS by~$\sim$~17--25\%.

The RELDIS model is based on a photonuclear reaction model~\cite{Ilinov1996,Pshenichnov2005}, which simulates the interaction of a photon with a nucleus as a four-step process. The modeling begins with (1) an intranuclear cascade of particles induced by the photon inside the nucleus, followed by (2) pre-equilibrium emission of nucleons, followed in some events by (3) nucleon coalescence, and finally (4) the excited residual nucleus evaporates nucleons or undergoes fission.  RELDIS modeling with steps (1)--(4) has been validated with several sets of experimental data~\cite{Scheidenberger2004,Pshenichnov2005,Pshenichnov2011,ALICE2023n}. However, the multiplicity of neutrons and protons from EMD is sensitive to the calculated excitation energy of residual nuclei, and it is affected by the presence of nucleon coalescence, which in particular leads to the production of deuterons from pairs of free neutrons and protons.  Therefore, multiplicities of neutrons and protons are generally higher in RELDIS calculations in which only (1) cascades and (4) evaporation-fission are included.

Two kinds of RELDIS results for modeling with steps (1)--(4) (standard) and only with (1) and (4) (without pre-equilibrium and coalescence) are presented in Table~\ref{tab:p_cs} allowing an assessment of the level of uncertainties of the calculations. Since in the latter case the excitation energy of the residual nuclei before nucleon evaporation is higher, the calculated cross sections of 1p, 2p, and 3p emission are also larger by 7\%, 13\%, and 21\%, respectively. In contrast, the calculated 0p cross section is smaller.

\begin{table}[!tb]
\caption{The cross sections of the emission of zero, one, two, and three protons accompanied by at least one neutron measured on the C and A sides. Cross sections calculated with RELDIS are given for comparison. The sum of 0p--3p cross sections is presented in the last row.  The uncertainties are given as $\pm$ (stat.) $\pm$ (syst.).}
\label{tab:p_cs}
\begin{center}
\begin{tabular}{|c|l|l|l|c|c|}
\hline
 & \multicolumn{2}{c|}{} & \multicolumn{1}{c|}{}& \multicolumn{2}{c|}{}\\
 & \multicolumn{2}{c|}{$\sigma(k{\mathrm{p}}, \mathrm{Xn})$ (b)}& \multicolumn{1}{c|}{$\sigma(k{\mathrm{p}}, \mathrm{Xn})$ (b)}& \multicolumn{2}{c|}{$\sigma^{\mathrm{RELDIS}}(k{\mathrm{p}}, \mathrm{Xn})$ (b)} \\
$k$p  & \multicolumn{2}{c|}{} & \multicolumn{1}{c|}{}& \multicolumn{2}{c|}{}\\
\cline{2-3} \cline{5-6}
& \multicolumn{1}{c|}{Side C} & \multicolumn{1}{c|}{Side A} & & \multicolumn{1}{c|}{std.} & \multicolumn{1}{c|}{w/o pre-eq.} \\
\hline
0p &  $ 156.5  \pm 0.2 \pm  5.4 $  & $ 158.5  \pm 0.2 \pm 5.5 $  & $ 157.5 \pm 0.1 \pm 4.7  $  & $155.5 $ & $147.0$ \\
\hline
1p &  $ 41.8 \pm  0.1 \pm 1.9 $  & $ 39.1 \pm 0.1  \pm 2.2 $  & $ 40.4 \pm 0.1  \pm 1.6 $  & $ 31.6 $ & $33.7$ \\
\hline
2p &  $ 17.4 \pm 0.2  \pm 5.9 $ & $ 16.3 \pm  0.2 \pm 4.3 $   & $ 16.8 \pm 0.1  \pm 3.7 $   & $ 11.2 $ & $12.7$ \\
\hline
3p &  $ 7.8 \pm 0. 5 \pm 2.4  $  & $ 5.9 \pm 0.4  \pm 3.6 $    & $ 6.8 \pm 0.3  \pm 2.2 $  & $ 5.3 $ & $ 6.4$ \\
\hline
0p--3p &  &   & $ 221.5 \pm 0.4  \pm 7.5 $  & $ 203.5 $ & $199.7$ \\
\hline
\end{tabular}
\end{center}
\end{table}

According to the Weizsacker--Williams method, the emission of neutrons and protons in EMD is induced mostly by absorbing low-energy equivalent photons. Because of this a single heavy residual nucleus along with few free nucleons is a dominant output of low-energy photonuclear reactions. In Ref.~\cite{ALICE2023n} the RELDIS results for the difference $\Delta A$ ($\Delta Z$) between the sum of the mass (charge) number of the most heavy residual nucleus and the number of emitted nucleons (protons) in each EMD event and $A$ ($Z$) of the initial $^{208}$Pb were presented.  The values of $\Delta A=0$, $\Delta Z=0$ were calculated for the majority of 1n--5n events.

Regarding the proton emission in EMD, the assumption of the dominance of the production of a single secondary nucleus also suggests that the emission of one, two, and three protons can be associated in general with the production of the respective elements: Tl, Hg, and Au. In order to illustrate this, the  cross sections of the production of various elements, Tl, Hg, Au, Pt, Ir, Os, Re, W, Ta, and Hf, were calculated using RELDIS in its standard mode. These cross sections are also plotted in Fig.~\ref{fig:protons} (dashed black line). As can be seen, the calculated production cross sections of certain elements are equal or above the calculated emission cross sections of the corresponding number of protons $k=0,1,...10$. The difference is noticeable for Pt, Ir, Os, Re, W, Ta, and Hf because their production by energetic photons is frequently accompanied by the emission of charged pions, deuterons, or alpha particles in addition to protons, see Ref.~\cite{ALICE2023n} for details.

The validity of RELDIS in calculating the production of Tl, Hg and Au as secondary nuclei in EMD associated, respectively, with the 1p, 2p and 3p emission studied in the present work, can be demonstrated by comparison with the charge-changing cross sections measured at lower energies for these elements at the CERN SPS~\cite{Scheidenberger2004,Cecchini2002}. Since those measurements contain contributions from both hadronic and ultraperipheral collisions, it is necessary to subtract the former from the data ~\cite{Scheidenberger2004,Cecchini2002} for comparison with the present ALICE data and RELDIS calculations, which include only UPCs. The hadronic fragmentation cross sections were estimated with the AAMCC-MST  model~\cite{Nepeivoda2022,Kozyrev2022}, and the results of their subtraction are also presented in Fig.~\ref{fig:protons}. The hadronic contributions to the measured cross sections to produce Tl in Pb--Pb and Pb--Au collisions at $\sqrt{s_{\mathrm{NN}}}=17.21$~GeV were calculated as 10\% and 18\%, respectively. Larger hadronic contributions of 24--39\% were calculated for the production of Hg and Au. As can be seen, the EMD cross sections estimated from the measurements at the CERN SPS~\cite{Scheidenberger2004,Cecchini2002} are in a good agreement with the corresponding cross sections calculated with RELDIS for the production of Tl, Hg, Au, and Pt. However, the differences between measurements and calculations are large for production of other elements, in particular for Ir, Ta, and Hf, where the dominant physics processes are different from those for Tl, Hg, Au, and Pt.  It can be also seen that the results of two experiments diverge for $k>4$, indicating potentially large uncertainties in both measurements. Nevertheless, it can be concluded that the EMD cross sections specifically for the production of Tl, Hg, and Au measured at both collision energies, $\sqrt{s_{\mathrm{NN}}}=17.21$~GeV and $\sqrt{s_{\mathrm{NN}}}=5.02$~TeV, are in agreement with theoretical results.   

It is interesting to note that the 3p emission cross section corresponding to the production of gold nuclei is measured to be $6.8\pm 2.2$~b.  Thus, the total inelastic hadronic cross section of $^{208}$Pb--$^{208}$Pb collisions ($7.67\pm 0.25$~b~\cite{Castellanos2021}) at the LHC is comparable to the cross section to produce a gold nucleus in the EMD of $^{208}$Pb. This means that hadronic $^{208}$Pb--$^{208}$Pb collisions, the main subject of heavy-ion physics at the LHC, occur almost as often as the production of undetected and mostly unstable gold nuclei from the lead nuclei of the colliding beams. However, the gold isotopes produced by the emission of 3 to 7 neutrons would be intercepted by collimators along the beamlines and would fragment, creating hadronic showers, on impact. Others are expected to be lost in the beam vacuum chamber in locations between the regions near the experiments where the beams are separated horizontally and the beginning of the regular collider arcs.

\subsection{Emission of protons associated with the production of \texorpdfstring{$^{206,205,204}$Tl}{206,205,204 Tl}}
\label{Sec:n_1p_mult}

\begin{figure}[!tb]
\begin{centering}
\includegraphics[width=0.9\columnwidth]{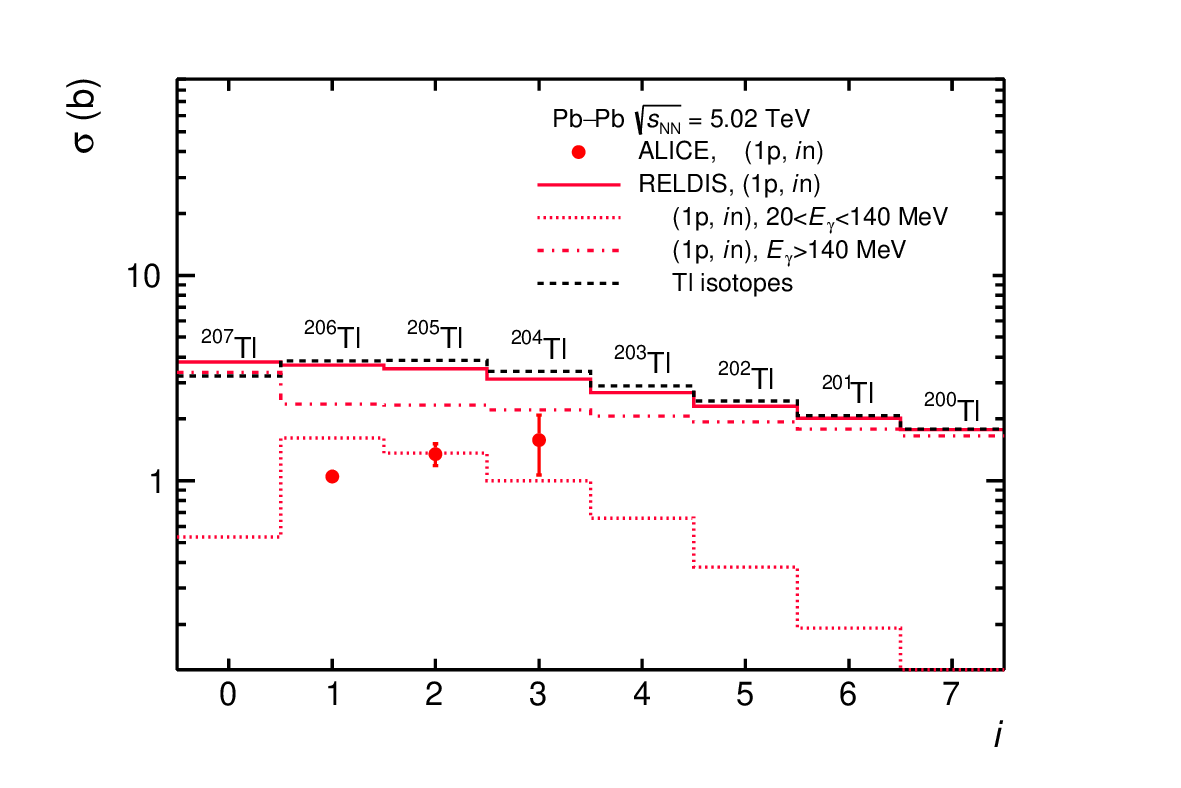}
\caption{Measured (circles) and calculated with RELDIS (solid-line histogram) cross sections of the emission of a given number of neutrons $i$ accompanied by a single proton in UPCs of $^{208}$Pb nuclei at $\sqrt{s_{\mathrm{NN}}}=5.02$~TeV. Error bars represent combined statistical and systematic uncertainties of the measurements. Dotted-line and dashed-dotted-line histograms present the same cross sections, but calculated with RELDIS for EMD induced by photons with energies in the domain of the quasideuteron absorption and hadron photoproduction on individual nucleons, respectively. Calculated cross sections of the production of specific thallium nuclei, $^{207,206,...,200}$Tl, are presented by the dashed-line histogram marked with the respective nuclide symbols.}  
\label{fig:Tl}
\end{centering}
\end{figure}

The measured (1p,1n), (1p,2n), and (1p,3n) cross sections are presented in Table~\ref{tab:n_p_cs} along with the respective cross sections calculated with RELDIS in its standard mode and also without the pre-equilibrium and coalescence contributions. The (1p,1n), (1p,2n), and (1p,3n) cross sections are associated with the production of specific thallium isotopes, $^{206,205,204}$Tl. Therefore, both kinds of calculated cross sections are presented in Fig.~\ref{fig:Tl} along with the data (red circles). The production cross sections of other thallium nuclei were also calculated with RELDIS in its standard mode and are also shown in Fig.~\ref{fig:Tl} along with the cross sections of the emission of corresponding numbers of neutrons along with a single proton. As can be seen, the calculated cross sections of the production of specific thallium isotopes (dashed black line) are very close to the calculated cross sections of the emission of the corresponding number of neutrons $i=0,1,...7$ accompanied by the 1p emission (solid red line). 
\begin{table}[!b]
\caption{Cross sections of the 1p emission accompanied by one, two, and three neutrons  measured on the C and A sides. Cross sections calculated with RELDIS are given for comparison. The uncertainties are given as $\pm$ (stat.) $\pm$ (syst.). }
\begin{center}
\begin{tabular}{|c|c|c|c|c|c|c|}
\hline
  &  & \multicolumn{2}{c|}{}&  & \multicolumn{2}{c|}{}  \\
 $k$p & $i$n & \multicolumn{2}{c|}{ $\sigma({1\mathrm{p},i\mathrm{n}})$ (b)}& { $\sigma({1\mathrm{p},i\mathrm{n}})$ (b)} & \multicolumn{2}{c|}{$\sigma^{\rm RELDIS}({1\mathrm{p},i\mathrm{n}})$ (b) }  \\
   &  & \multicolumn{2}{c|}{}&  & \multicolumn{2}{c|}{}  \\
\cline{3-4} \cline{6-7}
& & Side C & Side A & & std. & w/o pre-eq. \\
\hline
1p & 1n & $  1.01 \pm  0.02 \pm 0.04 $  & $  1.11 \pm 0.03 \pm 0.05 $  &  $ 1.05  \pm 0.03 \pm 0.04$   & $  3.67  $ & $3.95$ \\
\hline
1p & 2n & $1.30 \pm  0.07 \pm  0.13  $  & $ 1.25 \pm 0.06 \pm 0.09  $  &$ 1.35 \pm 0.05 \pm 0.16 $ & $ 3.52 $ & $3.86 $\\
\hline
1p & 3n & $ 1.32 \pm 0.10  \pm 0.12 $  & $  1.73 \pm 0.04 \pm 0.86  $  &  $ 1.58 \pm 0.06 \pm 0.51 $ & $ 3.12 $ & $3.27$\\
\hline
\end{tabular}
\end{center}
\label{tab:n_p_cs}
\end{table}

While most of the neutrons are emitted following the excitation of $^{208}$Pb by low-energy photons in the region of the GDR, it can be expected that a major part of $^{206,205,204}$Tl is produced due to the absorption of photons of higher energy $E_\gamma > 20 $~MeV~\cite{Dmitrieva2023}.  In fact, it is the region of $E_\gamma$ which lies above the proton emission threshold where the dominant processes are the absorption of photons on quasideuterons and the hadron photoproduction on individual nucleons.  This is supported by calculations with RELDIS of the contributions to the (1p,1n), (1p,2n), and (1p,3n) emission cross sections from photons of $20<E_\gamma<140$~MeV and $E_\gamma>140$~MeV, as also shown in Fig.~\ref{fig:Tl} (dotted red line and dashed-dotted red line, respectively). According to these calculations, the contribution by photons of $E_\gamma>140$~MeV to the emission of a single proton is the largest for all neutron multiplicities. 

As can be seen from Table~\ref{tab:n_p_cs} and Fig.~\ref{fig:Tl}, the measured (1p,1n), (1p,2n), and (1p,3n) cross sections are overestimated by RELDIS by a factor of 2--3. To the best of our knowledge, the data on the single proton emission following photoabsorption on $^{208}$Pb are very scarce. The corresponding cross sections were measured 40--60 years ago, either by detecting the activity of the resulting $^{206}$Tl and $^{207}$Tl in targets of $^{208}$Pb/$^\mathrm{nat}$Pb~\cite{Cameron1951,Sorokin1963,Dahmen1971}, or by the direct detection of the emitted photoprotons~\cite{Shoda1979}. While the cross sections measured long time ago differ by a factor of 2 or 3, the proton emission cross section calculated by RELDIS is closer to the earlier ones~\cite{Cameron1951,Sorokin1963}, but overestimates the results of two other experiments~\cite{Dahmen1971,Shoda1979} by a factor of 2--3. 

It should be noted, that the values of the total photoabsorption cross section on $^{208}$Pb measured in different experiments~\cite{Belyaev1992, Veyssiere1970, Harvey1964,Lepretre1981,Carlos1984} are characterized by large measurement uncertainties and diverge significantly for $20<E_\gamma<140$~MeV where the 1p emission from $^{208}$Pb is essential. A phenomenological approximation is presently used in RELDIS to describe the average of existing data on the total photoabsorption cross section on $^{208}$Pb.

Thus, based on the comparison of the available photoabsorption data with calculations, a possible overestimate of the 1p emission by RELDIS by a factor of 2--3 cannot be ruled out. This  may  explain a similar overestimate observed for the EMD 1p cross sections measured in the present work, see Fig.~\ref{fig:Tl}.

Although the noticeable overestimation of the (1p,1n), (1p,2n), and (1p,3n) cross sections by RELDIS seems contradictory to the underestimation of the (1p,Xn) cross section, it can be explained by the different multiplicity trends of the partial cross sections in model and data. It can be seen from Table~\ref{tab:n_p_cs} that RELDIS predicts a decrease of the cross sections with neutron multiplicity. In contrast, the measured cross sections increase with multiplicity. Hence, the unmeasured cross sections of the emission of more than three neutrons along with a single proton essentially contribute to the (1p,Xn) cross section and alleviate the difference between RELDIS and data for (1p,Xn). As a result, with a large discrepancy between calculations and data for (1p,1n), (1p,2n), and (1p,3n) seen in Fig.~\ref{fig:Tl} a better agreement is seen in Fig.~\ref{fig:protons} for (1p,Xn).

\section{Conclusions and outlook}\label{Sec:Conclusions}

The cross sections of the (0p,Xn),  (1p,Xn), (2p,Xn), and (3p,Xn) emission, as well as the cross sections of (1p,1n), (1p,2n), and (1p,3n) channels in EMD of $^{208}$Pb in UPCs were measured for the first time.  This study challenges models that aim to describe proton emission in the EMD of $^{208}$Pb at LHC energies. 

The measured cross sections for certain channels such as (1p,1n), (1p,2n), and (1p,3n) are significantly overestimated by RELDIS. For the more frequent events where at least one neutron is emitted, the measured 0p and 3p cross sections are described by RELDIS, while the 1p and 2p cross sections are underestimated by the model by~$\sim$~17--25\%. It is  shown that the calculated 0p, 1p, 2p, and 3p cross sections are sensitive to the involvement of pre-equilibrium emission and nucleon coalescence. In general, these measurements provide new information on photonuclear reactions and can be very useful in tuning models of photonuclear reactions as well as EMD models.

According to RELDIS, the 0p, 1p, 2p, and 3p cross sections are well suited to approximate the cross sections of production of certain elements, Pb, Tl, Hg, and Au, as secondary nuclei in EMD. In this context, the measured cross sections can be used as approximations of the cross sections of the production of Pb, Tl, Hg, and Au isotopes at the LHC.

It is known that gold isotopes are produced in nuclear reactions induced by low-energy neutrons, protons, deuterons and photons on Hg, Au and Pt targets (see~\cite{Kazakov2022} for a review). Gold isotopes were also produced in the fragmentation of target $^{209}$Bi nuclei induced by high-energy beams of $^{12}$C and $^{20}$Ne~\cite{Aleklett1981}, and in the fragmentation of uranium nuclei induced by high-energy protons~\cite{Barzakh2022}. However, according to the Experimental Nuclear Reaction Data (EXFOR) database~\cite{Zerkin2018}, no data were reported specifically for the photonuclear reactions Pb$^*$($\gamma$, $^*$)Au$^*$. In the present work we report for the first time the production of gold in the photoabsorption by lead nuclei, purely in ultraperipheral $^{208}$Pb--$^{208}$Pb  collisions, with unusually large cross section comparable to the total hadronic $^{208}$Pb--$^{208}$Pb cross section. It can be estimated that a total of some $2.9\times10^{-11}$~g of various gold isotopes were produced from both LHC beams according to the total integrated luminosity at all four interaction points during Run 2 (2015--2018) of the LHC. The Pb--Pb data analyzed in the present paper were collected in 2018. The transmutation of lead into gold is the dream of medieval alchemists which comes true at the LHC.

Although only fluxes of $^{206}$Pb and $^{207}$Pb from EMD that accompany the $^{208}$Pb beam and are not intercepted by collimators pose a risk to LHC operation~\cite{Bruce2009,Klein2000}, the production of other secondary nuclei still significantly limits beam lifetime and reduces luminosity.  Thus, these new measurements provide also a valuable contribution to the study of beam losses due to electromagnetic processes at the LHC and future colliders.


\newenvironment{acknowledgement}{\relax}{\relax}
\begin{acknowledgement}
\section*{Acknowledgements}
\input{fa_2024-10-10_Opt_C.tex}
\end{acknowledgement}

\bibliographystyle{utphys}   
\bibliography{ProtonsPbPaper}

\newpage
\appendix

%
%

\section{The ALICE Collaboration}
\label{app:collab}
\input{Alice_Authorlist_2024-10-10_Opt_C.tex}
\end{document}

%% file: commands.tex
%

\newcommand{\pp}           {pp\xspace}
\newcommand{\ppbar}        {\mbox{$\mathrm {p\overline{p}}$}\xspace}
\newcommand{\XeXe}         {\mbox{Xe--Xe}\xspace}
\newcommand{\PbPb}         {\mbox{Pb--Pb}\xspace}
\newcommand{\pA}           {\mbox{pA}\xspace}
\newcommand{\pPb}          {\mbox{p--Pb}\xspace}
\newcommand{\AuAu}         {\mbox{Au--Au}\xspace}
\newcommand{\dAu}          {\mbox{d--Au}\xspace}

\newcommand{\s}            {\ensuremath{\sqrt{s}}\xspace}
\newcommand{\snn}          {\ensuremath{\sqrt{s_{\mathrm{NN}}}}\xspace}
\newcommand{\pt}           {\ensuremath{p_{\rm T}}\xspace}
\newcommand{\meanpt}       {$\langle p_{\mathrm{T}}\rangle$\xspace}
\newcommand{\ycms}         {\ensuremath{y_{\rm CMS}}\xspace}
\newcommand{\ylab}         {\ensuremath{y_{\rm lab}}\xspace}
\newcommand{\etarange}[1]  {\mbox{$\left | \eta \right |~<~#1$}}
\newcommand{\yrange}[1]    {\mbox{$\left | y \right |~<~#1$}}
\newcommand{\dndy}         {\ensuremath{\mathrm{d}N_\mathrm{ch}/\mathrm{d}y}\xspace}
\newcommand{\dndeta}       {\ensuremath{\mathrm{d}N_\mathrm{ch}/\mathrm{d}\eta}\xspace}
\newcommand{\avdndeta}     {\ensuremath{\langle\dndeta\rangle}\xspace}
\newcommand{\dNdy}         {\ensuremath{\mathrm{d}N_\mathrm{ch}/\mathrm{d}y}\xspace}
\newcommand{\Npart}        {\ensuremath{N_\mathrm{part}}\xspace}
\newcommand{\Ncoll}        {\ensuremath{N_\mathrm{coll}}\xspace}
\newcommand{\dEdx}         {\ensuremath{\textrm{d}E/\textrm{d}x}\xspace}
\newcommand{\RpPb}         {\ensuremath{R_{\rm pPb}}\xspace}

\newcommand{\nineH}        {$\sqrt{s}~=~0.9$~Te\kern-.1emV\xspace}
\newcommand{\seven}        {$\sqrt{s}~=~7$~Te\kern-.1emV\xspace}
\newcommand{\twoH}         {$\sqrt{s}~=~0.2$~Te\kern-.1emV\xspace}
\newcommand{\twosevensix}  {$\sqrt{s}~=~2.76$~Te\kern-.1emV\xspace}
\newcommand{\five}         {$\sqrt{s}~=~5.02$~Te\kern-.1emV\xspace}
\newcommand{\twosevensixnn}{$\sqrt{s_{\mathrm{NN}}}~=~2.76$~Te\kern-.1emV\xspace}
\newcommand{\fivenn}       {$\sqrt{s_{\mathrm{NN}}}~=~5.02$~Te\kern-.1emV\xspace}
\newcommand{\LT}           {L{\'e}vy-Tsallis\xspace}
\newcommand{\GeVc}         {Ge\kern-.1emV/$c$\xspace}
\newcommand{\MeVc}         {Me\kern-.1emV/$c$\xspace}
\newcommand{\TeV}          {Te\kern-.1emV\xspace}
\newcommand{\GeV}          {Ge\kern-.1emV\xspace}
\newcommand{\MeV}          {Me\kern-.1emV\xspace}
\newcommand{\GeVmass}      {Ge\kern-.2emV/$c^2$\xspace}
\newcommand{\MeVmass}      {Me\kern-.2emV/$c^2$\xspace}
\newcommand{\lumi}         {\ensuremath{\mathcal{L}}\xspace}

\newcommand{\ITS}          {\rm{ITS}\xspace}
\newcommand{\TOF}          {\rm{TOF}\xspace}
\newcommand{\ZDC}          {\rm{ZDC}\xspace}
\newcommand{\ZDCs}         {\rm{ZDCs}\xspace}
\newcommand{\ZNA}          {\rm{ZNA}\xspace}
\newcommand{\ZNC}          {\rm{ZNC}\xspace}
\newcommand{\SPD}          {\rm{SPD}\xspace}
\newcommand{\SDD}          {\rm{SDD}\xspace}
\newcommand{\SSD}          {\rm{SSD}\xspace}
\newcommand{\TPC}          {\rm{TPC}\xspace}
\newcommand{\TRD}          {\rm{TRD}\xspace}
\newcommand{\VZERO}        {\rm{V0}\xspace}
\newcommand{\VZEROA}       {\rm{V0A}\xspace}
\newcommand{\VZEROC}       {\rm{V0C}\xspace}
\newcommand{\Vdecay} 	   {\ensuremath{V^{0}}\xspace}

\newcommand{\ee}           {\ensuremath{e^{+}e^{-}}} 
\newcommand{\pip}          {\ensuremath{\pi^{+}}\xspace}
\newcommand{\pim}          {\ensuremath{\pi^{-}}\xspace}
\newcommand{\kap}          {\ensuremath{\rm{K}^{+}}\xspace}
\newcommand{\kam}          {\ensuremath{\rm{K}^{-}}\xspace}
\newcommand{\pbar}         {\ensuremath{\rm\overline{p}}\xspace}
\newcommand{\kzero}        {\ensuremath{{\rm K}^{0}_{\rm{S}}}\xspace}
\newcommand{\lmb}          {\ensuremath{\Lambda}\xspace}
\newcommand{\almb}         {\ensuremath{\overline{\Lambda}}\xspace}
\newcommand{\Om}           {\ensuremath{\Omega^-}\xspace}
\newcommand{\Mo}           {\ensuremath{\overline{\Omega}^+}\xspace}
\newcommand{\X}            {\ensuremath{\Xi^-}\xspace}
\newcommand{\Ix}           {\ensuremath{\overline{\Xi}^+}\xspace}
\newcommand{\Xis}          {\ensuremath{\Xi^{\pm}}\xspace}
\newcommand{\Oms}          {\ensuremath{\Omega^{\pm}}\xspace}
\newcommand{\degree}       {\ensuremath{^{\rm o}}\xspace}

%% file: fa_2024-10-10_Opt_C.tex

The ALICE Collaboration would like to thank all its engineers and technicians for their invaluable contributions to the construction of the experiment and the CERN accelerator teams for the outstanding performance of the LHC complex.
The ALICE Collaboration gratefully acknowledges the resources and support provided by all Grid centres and the Worldwide LHC Computing Grid (WLCG) collaboration.
The ALICE Collaboration acknowledges the following funding agencies for their support in building and running the ALICE detector:
A. I. Alikhanyan National Science Laboratory (Yerevan Physics Institute) Foundation (ANSL), State Committee of Science and World Federation of Scientists (WFS), Armenia;
Austrian Academy of Sciences, Austrian Science Fund (FWF): [M 2467-N36] and Nationalstiftung f\"{u}r Forschung, Technologie und Entwicklung, Austria;
Ministry of Communications and High Technologies, National Nuclear Research Center, Azerbaijan;
Conselho Nacional de Desenvolvimento Cient\'{\i}fico e Tecnol\'{o}gico (CNPq), Financiadora de Estudos e Projetos (Finep), Funda\c{c}\~{a}o de Amparo \`{a} Pesquisa do Estado de S\~{a}o Paulo (FAPESP) and Universidade Federal do Rio Grande do Sul (UFRGS), Brazil;
Bulgarian Ministry of Education and Science, within the National Roadmap for Research Infrastructures 2020-2027 (object CERN), Bulgaria;
Ministry of Education of China (MOEC) , Ministry of Science \& Technology of China (MSTC) and National Natural Science Foundation of China (NSFC), China;
Ministry of Science and Education and Croatian Science Foundation, Croatia;
Centro de Aplicaciones Tecnol\'{o}gicas y Desarrollo Nuclear (CEADEN), Cubaenerg\'{\i}a, Cuba;
Ministry of Education, Youth and Sports of the Czech Republic, Czech Republic;
The Danish Council for Independent Research | Natural Sciences, the VILLUM FONDEN and Danish National Research Foundation (DNRF), Denmark;
Helsinki Institute of Physics (HIP), Finland;
Commissariat \`{a} l'Energie Atomique (CEA) and Institut National de Physique Nucl\'{e}aire et de Physique des Particules (IN2P3) and Centre National de la Recherche Scientifique (CNRS), France;
Bundesministerium f\"{u}r Bildung und Forschung (BMBF) and GSI Helmholtzzentrum f\"{u}r Schwerionenforschung GmbH, Germany;
General Secretariat for Research and Technology, Ministry of Education, Research and Religions, Greece;
National Research, Development and Innovation Office, Hungary;
Department of Atomic Energy Government of India (DAE), Department of Science and Technology, Government of India (DST), University Grants Commission, Government of India (UGC) and Council of Scientific and Industrial Research (CSIR), India;
National Research and Innovation Agency - BRIN, Indonesia;
Istituto Nazionale di Fisica Nucleare (INFN), Italy;
Japanese Ministry of Education, Culture, Sports, Science and Technology (MEXT) and Japan Society for the Promotion of Science (JSPS) KAKENHI, Japan;
Consejo Nacional de Ciencia (CONACYT) y Tecnolog\'{i}a, through Fondo de Cooperaci\'{o}n Internacional en Ciencia y Tecnolog\'{i}a (FONCICYT) and Direcci\'{o}n General de Asuntos del Personal Academico (DGAPA), Mexico;
Nederlandse Organisatie voor Wetenschappelijk Onderzoek (NWO), Netherlands;
The Research Council of Norway, Norway;
Pontificia Universidad Cat\'{o}lica del Per\'{u}, Peru;
Ministry of Science and Higher Education, National Science Centre and WUT ID-UB, Poland;
Korea Institute of Science and Technology Information and National Research Foundation of Korea (NRF), Republic of Korea;
Ministry of Education and Scientific Research, Institute of Atomic Physics, Ministry of Research and Innovation and Institute of Atomic Physics and Universitatea Nationala de Stiinta si Tehnologie Politehnica Bucuresti, Romania;
Ministry of Education, Science, Research and Sport of the Slovak Republic, Slovakia;
National Research Foundation of South Africa, South Africa;
Swedish Research Council (VR) and Knut \& Alice Wallenberg Foundation (KAW), Sweden;
European Organization for Nuclear Research, Switzerland;
Suranaree University of Technology (SUT), National Science and Technology Development Agency (NSTDA) and National Science, Research and Innovation Fund (NSRF via PMU-B B05F650021), Thailand;
Turkish Energy, Nuclear and Mineral Research Agency (TENMAK), Turkey;
National Academy of  Sciences of Ukraine, Ukraine;
Science and Technology Facilities Council (STFC), United Kingdom;
National Science Foundation of the United States of America (NSF) and United States Department of Energy, Office of Nuclear Physics (DOE NP), United States of America.
In addition, individual groups or members have received support from:
Czech Science Foundation (grant no. 23-07499S), Czech Republic;
FORTE project, reg.\ no.\ CZ.02.01.01/00/22\_008/0004632, Czech Republic, co-funded by the European Union, Czech Republic;
European Research Council (grant no. 950692), European Union;
ICSC - Centro Nazionale di Ricerca in High Performance Computing, Big Data and Quantum Computing, European Union - NextGenerationEU;
Academy of Finland (Center of Excellence in Quark Matter) (grant nos. 346327, 346328), Finland;
Deutsche Forschungs Gemeinschaft (DFG, German Research Foundation) ``Neutrinos and Dark Matter in Astro- and Particle Physics'' (grant no. SFB 1258), Germany.

%% file: Alice_Authorlist_2024-10-10_Opt_C.tex
\begin{flushleft} 
\small

S.~Acharya\,\orcidlink{0000-0002-9213-5329}\,$^{\rm 126}$, 
A.~Agarwal$^{\rm 134}$, 
G.~Aglieri Rinella\,\orcidlink{0000-0002-9611-3696}\,$^{\rm 32}$, 
L.~Aglietta\,\orcidlink{0009-0003-0763-6802}\,$^{\rm 24}$, 
M.~Agnello\,\orcidlink{0000-0002-0760-5075}\,$^{\rm 29}$, 
N.~Agrawal\,\orcidlink{0000-0003-0348-9836}\,$^{\rm 25}$, 
Z.~Ahammed\,\orcidlink{0000-0001-5241-7412}\,$^{\rm 134}$, 
S.~Ahmad\,\orcidlink{0000-0003-0497-5705}\,$^{\rm 15}$, 
S.U.~Ahn\,\orcidlink{0000-0001-8847-489X}\,$^{\rm 71}$, 
I.~Ahuja\,\orcidlink{0000-0002-4417-1392}\,$^{\rm 36}$
% ...

\section*{Affiliation Notes}

$^{\rm I}$ Deceased\\
$^{\rm II}$ Also at: Max-Planck-Institut fur Physik, Munich, Germany\\
$^{\rm III}$ Also at: Italian National Agency for New Technologies, Energy and Sustainable Economic Development (ENEA), Bologna, Italy\\
$^{\rm IV}$ Also at: Dipartimento DET del Politecnico di Torino, Turin, Italy\\
$^{\rm V}$ Also at: Department of Applied Physics, Aligarh Muslim University, Aligarh, India\\
$^{\rm VI}$ Also at: Institute of Theoretical Physics, University of Wroclaw, Poland\\
$^{\rm VII}$ Also at: Facultad de Ciencias, Universidad Nacional Autónoma de México, Mexico City, Mexico\\

\section*{Collaboration Institutes}

$^{1}$ A.I. Alikhanyan National Science Laboratory (Yerevan Physics Institute) Foundation, Yerevan, Armenia\\
$^{2}$ AGH University of Krakow, Cracow, Poland\\
$^{3}$ Bogolyubov Institute for Theoretical Physics, National Academy of Sciences of Ukraine, Kiev, Ukraine\\
$^{4}$ Bose Institute, Department of Physics  and Centre for Astroparticle Physics and Space Science (CAPSS), Kolkata, India\\
$^{5}$ California Polytechnic State University, San Luis Obispo, California, United States\\
$^{6}$ Central China Normal University, Wuhan, China\\
$^{7}$ Centro de Aplicaciones Tecnol\'{o}gicas y Desarrollo Nuclear (CEADEN), Havana, Cuba\\
$^{8}$ Centro de Investigaci\'{o}n y de Estudios Avanzados (CINVESTAV), Mexico City and M\'{e}rida, Mexico\\
$^{9}$ Chicago State University, Chicago, Illinois, United States\\
$^{10}$ China Institute of Atomic Energy, Beijing, China\\
$^{11}$ China University of Geosciences, Wuhan, China\\
$^{12}$ Chungbuk National University, Cheongju, Republic of Korea\\
$^{13}$ Comenius University Bratislava, Faculty of Mathematics, Physics and Informatics, Bratislava, Slovak Republic\\
$^{14}$ Creighton University, Omaha, Nebraska, United States\\
$^{15}$ Department of Physics, Aligarh Muslim University, Aligarh, India\\
$^{16}$ Department of Physics, Pusan National University, Pusan, Republic of Korea\\
$^{17}$ Department of Physics, Sejong University, Seoul, Republic of Korea\\
$^{18}$ Department of Physics, University of California, Berkeley, California, United States\\
$^{19}$ Department of Physics, University of Oslo, Oslo, Norway\\
$^{20}$ Department of Physics and Technology, University of Bergen, Bergen, Norway\\
$^{21}$ Dipartimento di Fisica, Universit\`{a} di Pavia, Pavia, Italy\\
$^{22}$ Dipartimento di Fisica dell'Universit\`{a} and Sezione INFN, Cagliari, Italy\\
$^{23}$ Dipartimento di Fisica dell'Universit\`{a} and Sezione INFN, Trieste, Italy\\
$^{24}$ Dipartimento di Fisica dell'Universit\`{a} and Sezione INFN, Turin, Italy\\
$^{25}$ Dipartimento di Fisica e Astronomia dell'Universit\`{a} and Sezione INFN, Bologna, Italy\\
$^{26}$ Dipartimento di Fisica e Astronomia dell'Universit\`{a} and Sezione INFN, Catania, Italy\\
$^{27}$ Dipartimento di Fisica e Astronomia dell'Universit\`{a} and Sezione INFN, Padova, Italy\\
$^{28}$ Dipartimento di Fisica `E.R.~Caianiello' dell'Universit\`{a} and Gruppo Collegato INFN, Salerno, Italy\\
$^{29}$ Dipartimento DISAT del Politecnico and Sezione INFN, Turin, Italy\\
$^{30}$ Dipartimento di Scienze MIFT, Universit\`{a} di Messina, Messina, Italy\\
$^{31}$ Dipartimento Interateneo di Fisica `M.~Merlin' and Sezione INFN, Bari, Italy\\
$^{32}$ European Organization for Nuclear Research (CERN), Geneva, Switzerland\\
$^{33}$ Faculty of Electrical Engineering, Mechanical Engineering and Naval Architecture, University of Split, Split, Croatia\\
$^{34}$ Faculty of Nuclear Sciences and Physical Engineering, Czech Technical University in Prague, Prague, Czech Republic\\
$^{35}$ Faculty of Physics, Sofia University, Sofia, Bulgaria\\
$^{36}$ Faculty of Science, P.J.~\v{S}af\'{a}rik University, Ko\v{s}ice, Slovak Republic\\
$^{37}$ Faculty of Technology, Environmental and Social Sciences, Bergen, Norway\\
$^{38}$ Frankfurt Institute for Advanced Studies, Johann Wolfgang Goethe-Universit\"{a}t Frankfurt, Frankfurt, Germany\\
$^{39}$ Fudan University, Shanghai, China\\
$^{40}$ Gangneung-Wonju National University, Gangneung, Republic of Korea\\
$^{41}$ Gauhati University, Department of Physics, Guwahati, India\\
$^{42}$ Helmholtz-Institut f\"{u}r Strahlen- und Kernphysik, Rheinische Friedrich-Wilhelms-Universit\"{a}t Bonn, Bonn, Germany\\
$^{43}$ Helsinki Institute of Physics (HIP), Helsinki, Finland\\
$^{44}$ High Energy Physics Group,  Universidad Aut\'{o}noma de Puebla, Puebla, Mexico\\
$^{45}$ Horia Hulubei National Institute of Physics and Nuclear Engineering, Bucharest, Romania\\
$^{46}$ HUN-REN Wigner Research Centre for Physics, Budapest, Hungary\\
$^{47}$ Indian Institute of Technology Bombay (IIT), Mumbai, India\\
$^{48}$ Indian Institute of Technology Indore, Indore, India\\
$^{49}$ INFN, Laboratori Nazionali di Frascati, Frascati, Italy\\
$^{50}$ INFN, Sezione di Bari, Bari, Italy\\
$^{51}$ INFN, Sezione di Bologna, Bologna, Italy\\
$^{52}$ INFN, Sezione di Cagliari, Cagliari, Italy\\
$^{53}$ INFN, Sezione di Catania, Catania, Italy\\
$^{54}$ INFN, Sezione di Padova, Padova, Italy\\
$^{55}$ INFN, Sezione di Pavia, Pavia, Italy\\
$^{56}$ INFN, Sezione di Torino, Turin, Italy\\
$^{57}$ INFN, Sezione di Trieste, Trieste, Italy\\
$^{58}$ Inha University, Incheon, Republic of Korea\\
$^{59}$ Institute for Gravitational and Subatomic Physics (GRASP), Utrecht University/Nikhef, Utrecht, Netherlands\\
$^{60}$ Institute of Experimental Physics, Slovak Academy of Sciences, Ko\v{s}ice, Slovak Republic\\
$^{61}$ Institute of Physics, Homi Bhabha National Institute, Bhubaneswar, India\\
$^{62}$ Institute of Physics of the Czech Academy of Sciences, Prague, Czech Republic\\
$^{63}$ Institute of Space Science (ISS), Bucharest, Romania\\
$^{64}$ Institut f\"{u}r Kernphysik, Johann Wolfgang Goethe-Universit\"{a}t Frankfurt, Frankfurt, Germany\\
$^{65}$ Instituto de Ciencias Nucleares, Universidad Nacional Aut\'{o}noma de M\'{e}xico, Mexico City, Mexico\\
$^{66}$ Instituto de F\'{i}sica, Universidade Federal do Rio Grande do Sul (UFRGS), Porto Alegre, Brazil\\
$^{67}$ Instituto de F\'{\i}sica, Universidad Nacional Aut\'{o}noma de M\'{e}xico, Mexico City, Mexico\\
$^{68}$ iThemba LABS, National Research Foundation, Somerset West, South Africa\\
$^{69}$ Jeonbuk National University, Jeonju, Republic of Korea\\
$^{70}$ Johann-Wolfgang-Goethe Universit\"{a}t Frankfurt Institut f\"{u}r Informatik, Fachbereich Informatik und Mathematik, Frankfurt, Germany\\
$^{71}$ Korea Institute of Science and Technology Information, Daejeon, Republic of Korea\\
$^{72}$ Laboratoire de Physique Subatomique et de Cosmologie, Universit\'{e} Grenoble-Alpes, CNRS-IN2P3, Grenoble, France\\
$^{73}$ Lawrence Berkeley National Laboratory, Berkeley, California, United States\\
$^{74}$ Lund University Department of Physics, Division of Particle Physics, Lund, Sweden\\
$^{75}$ Nagasaki Institute of Applied Science, Nagasaki, Japan\\
$^{76}$ Nara Women{'}s University (NWU), Nara, Japan\\
$^{77}$ National and Kapodistrian University of Athens, School of Science, Department of Physics , Athens, Greece\\
$^{78}$ National Centre for Nuclear Research, Warsaw, Poland\\
$^{79}$ National Institute of Science Education and Research, Homi Bhabha National Institute, Jatni, India\\
$^{80}$ National Nuclear Research Center, Baku, Azerbaijan\\
$^{81}$ National Research and Innovation Agency - BRIN, Jakarta, Indonesia\\
$^{82}$ Niels Bohr Institute, University of Copenhagen, Copenhagen, Denmark\\
$^{83}$ Nikhef, National institute for subatomic physics, Amsterdam, Netherlands\\
$^{84}$ Nuclear Physics Group, STFC Daresbury Laboratory, Daresbury, United Kingdom\\
$^{85}$ Nuclear Physics Institute of the Czech Academy of Sciences, Husinec-\v{R}e\v{z}, Czech Republic\\
$^{86}$ Oak Ridge National Laboratory, Oak Ridge, Tennessee, United States\\
$^{87}$ Ohio State University, Columbus, Ohio, United States\\
$^{88}$ Physics department, Faculty of science, University of Zagreb, Zagreb, Croatia\\
$^{89}$ Physics Department, Panjab University, Chandigarh, India\\
$^{90}$ Physics Department, University of Jammu, Jammu, India\\
$^{91}$ Physics Program and International Institute for Sustainability with Knotted Chiral Meta Matter (WPI-SKCM$^{2}$), Hiroshima University, Hiroshima, Japan\\
$^{92}$ Physikalisches Institut, Eberhard-Karls-Universit\"{a}t T\"{u}bingen, T\"{u}bingen, Germany\\
$^{93}$ Physikalisches Institut, Ruprecht-Karls-Universit\"{a}t Heidelberg, Heidelberg, Germany\\
$^{94}$ Physik Department, Technische Universit\"{a}t M\"{u}nchen, Munich, Germany\\
$^{95}$ Politecnico di Bari and Sezione INFN, Bari, Italy\\
$^{96}$ Research Division and ExtreMe Matter Institute EMMI, GSI Helmholtzzentrum f\"ur Schwerionenforschung GmbH, Darmstadt, Germany\\
$^{97}$ Saga University, Saga, Japan\\
$^{98}$ Saha Institute of Nuclear Physics, Homi Bhabha National Institute, Kolkata, India\\
$^{99}$ School of Physics and Astronomy, University of Birmingham, Birmingham, United Kingdom\\
$^{100}$ Secci\'{o}n F\'{\i}sica, Departamento de Ciencias, Pontificia Universidad Cat\'{o}lica del Per\'{u}, Lima, Peru\\
$^{101}$ Stefan Meyer Institut f\"{u}r Subatomare Physik (SMI), Vienna, Austria\\
$^{102}$ SUBATECH, IMT Atlantique, Nantes Universit\'{e}, CNRS-IN2P3, Nantes, France\\
$^{103}$ Sungkyunkwan University, Suwon City, Republic of Korea\\
$^{104}$ Suranaree University of Technology, Nakhon Ratchasima, Thailand\\
$^{105}$ Technical University of Ko\v{s}ice, Ko\v{s}ice, Slovak Republic\\
$^{106}$ The Henryk Niewodniczanski Institute of Nuclear Physics, Polish Academy of Sciences, Cracow, Poland\\
$^{107}$ The University of Texas at Austin, Austin, Texas, United States\\
$^{108}$ Universidad Aut\'{o}noma de Sinaloa, Culiac\'{a}n, Mexico\\
$^{109}$ Universidade de S\~{a}o Paulo (USP), S\~{a}o Paulo, Brazil\\
$^{110}$ Universidade Estadual de Campinas (UNICAMP), Campinas, Brazil\\
$^{111}$ Universidade Federal do ABC, Santo Andre, Brazil\\
$^{112}$ Universitatea Nationala de Stiinta si Tehnologie Politehnica Bucuresti, Bucharest, Romania\\
$^{113}$ University of Cape Town, Cape Town, South Africa\\
$^{114}$ University of Derby, Derby, United Kingdom\\
$^{115}$ University of Houston, Houston, Texas, United States\\
$^{116}$ University of Jyv\"{a}skyl\"{a}, Jyv\"{a}skyl\"{a}, Finland\\
$^{117}$ University of Kansas, Lawrence, Kansas, United States\\
$^{118}$ University of Liverpool, Liverpool, United Kingdom\\
$^{119}$ University of Science and Technology of China, Hefei, China\\
$^{120}$ University of South-Eastern Norway, Kongsberg, Norway\\
$^{121}$ University of Tennessee, Knoxville, Tennessee, United States\\
$^{122}$ University of the Witwatersrand, Johannesburg, South Africa\\
$^{123}$ University of Tokyo, Tokyo, Japan\\
$^{124}$ University of Tsukuba, Tsukuba, Japan\\
$^{125}$ Universit\"{a}t M\"{u}nster, Institut f\"{u}r Kernphysik, M\"{u}nster, Germany\\
$^{126}$ Universit\'{e} Clermont Auvergne, CNRS/IN2P3, LPC, Clermont-Ferrand, France\\
$^{127}$ Universit\'{e} de Lyon, CNRS/IN2P3, Institut de Physique des 2 Infinis de Lyon, Lyon, France\\
$^{128}$ Universit\'{e} de Strasbourg, CNRS, IPHC UMR 7178, F-67000 Strasbourg, France, Strasbourg, France\\
$^{129}$ Universit\'{e} Paris-Saclay, Centre d'Etudes de Saclay (CEA), IRFU, D\'{e}partment de Physique Nucl\'{e}aire (DPhN), Saclay, France\\
$^{130}$ Universit\'{e}  Paris-Saclay, CNRS/IN2P3, IJCLab, Orsay, France\\
$^{131}$ Universit\`{a} degli Studi di Foggia, Foggia, Italy\\
$^{132}$ Universit\`{a} del Piemonte Orientale, Vercelli, Italy\\
$^{133}$ Universit\`{a} di Brescia, Brescia, Italy\\
$^{134}$ Variable Energy Cyclotron Centre, Homi Bhabha National Institute, Kolkata, India\\
$^{135}$ Warsaw University of Technology, Warsaw, Poland\\
$^{136}$ Wayne State University, Detroit, Michigan, United States\\
$^{137}$ Yale University, New Haven, Connecticut, United States\\
$^{138}$ Yildiz Technical University, Istanbul, Turkey\\
$^{139}$ Yonsei University, Seoul, Republic of Korea\\
$^{140}$ Affiliated with an institute covered by a cooperation agreement with CERN\\
$^{141}$ Affiliated with an international laboratory covered by a cooperation agreement with CERN.\\

\end{flushleft} 